\documentstyle[epsfig]{mn2e}

\title[The impact of reionization on black hole formation]{The impact of
  reionization on the formation of supermassive black hole seeds} 

\author[J. L. Johnson, et al.]{Jarrett L. Johnson$^1$\thanks{Email:
    jlj@lanl.gov}, Daniel J. Whalen$^2$, Bhaskar Agarwal$^3$, Jan-Pieter 
\newauthor Paardekooper$^3$, and Sadegh Khochfar$^{4}$ \\
$^1$X Theoretical Division, Los Alamos National Laboratory, Los Alamos, NM  87545, USA \\
$^2$Universit{\"a}t Heidelberg, Zentrum f{\"u}r Astronomie, Institut
f{\"u}r Theoretische Astrophysik, \\
Albert-Ueberle-Str. 2, D-69120  Heidelberg, Germany \\
$^3$Max-Planck-Institut f{\"u}r extraterrestrische Physik,
Giessenbachstra\ss{}e, 85748 Garching, Germany\\
$^4$Institute for Astronomy, University of Edinburgh, Royal Observatory, Edinburgh, EH9 3HJ}

\pubyear{2014}

\begin{document}
\maketitle

\begin{abstract}
Direct collapse black holes (DCBHs) formed from the collapse of atomically-cooled primordial gas in the early Universe are strong candidates for the seeds of supermassive BHs.  DCBHs are thought to form in atomic cooling haloes in the presence of a strong molecule-dissociating, Lyman-Werner (LW) radiation field. Given that star forming galaxies are likely to be the source of the LW radiation in this scenario, ionizing radiation from these galaxies may accompany the LW radiation.  We present cosmological simulations resolving the collapse of primordial gas into an atomic cooling halo, including the effects of both LW and ionizing radiation.  We find that in cases where the gas is not self-shielded from the ionizing radiation, the collapse can be delayed by $\sim$ 25 Myr.  When the ionized gas does collapse, the free electrons that are present catalyze H$_{\rm 2}$ formation.  In turn, H$_{\rm 2}$ cooling becomes efficient in the center of the halo, and DCBH formation is prevented.  We emphasize, however, that in many cases the gas collapsing into atomic cooling haloes at high redshift is self-shielding to ionizing radiation.  Therefore, it is only in a fraction of such haloes in which DCBH formation is prevented due to reionization.  
\end{abstract}


\section{Introduction}
The origin of the black holes (BHs) inhabiting the centers of massive
galaxies (e.g. Gebhardt et al. 2000; Merritt \& Ferrarese 2001) and powering 
luminous quasars at high redshift 
(e.g. Willott et al. 2003; Fan et al. 2006; Mortlock et al. 2011) has long been an open question at 
the forefront of cosmology and galaxy formation.  There is a strong 
possibility that many of these grew from seed BHs which were born 
with masses of 10$^4$ - 10$^6$ M$_{\odot}$ in the centers of atomic cooling 
dark matter (DM) halos in the early universe (e.g. Volonteri 2012;
Natarajan \& Volonteri 2012). 
In this scenario, the primordial gas within the halo is unable
to cool below $\sim$ 10$^4$ K (e.g. Spaans \& Silk 2006) because of a low abundance of H$_{\rm 2}$ molecules, 
which implies that runaway gravitational collapse only occurs once 
up to $\sim$ 10$^6$ M$_{\odot}$ of gas has accumulated in the center of the halo.  
The central objects that form from this collapse, likely short-lived 
supermassive stars (e.g. Fuller et al. 1986; Hosokawa et al. 2013) or quasi-stars
(e.g. Begelman et al. 2008), grow at rates of up to 1 M$_{\odot}$
yr$^{-1}$ (Wise et al. 2008; Regan \& Haehnelt 2009; Shang et
al. 2010; Johnson et al. 2011; Latif et al. 2013a; Prieto et al. 2013) and are believed to typically leave behind BHs with
masses of up to 10$^6$ M$_{\odot}$ (e.g. Choi et al. 2013).\footnote{We note that a small fraction of supermassive stars may instead explode as powerful supernovae and leave behind no remnant (e.g. Montero et al. 2012; Chen et al. 2014).}  BHs formed via this process are 
referred to as direct collapse black holes (DCBHs).

In DCBH formation, the H$_{\rm 2}$ fraction is typically considered to be
suppressed by a strong molecule-dissociating, Lyman-Werner (LW)
radiation field (e.g. Haiman 2000; Glover \& Brand 2001; 
Machacek et al. 2001; Ahn et al. 2012).  While an elevated LW radiation
field is likely required to sufficiently suppress H$_{\rm 2}$ cooling during
the collapse of the gas (e.g. Bromm \& Loeb 2003; Shang et al. 2010; Van Borm
\& Spaans 2013; Visbal et al. 2014a), the results of both
semi-analytic models (e.g. Dijkstra et al. 2008; Petri et al. 2012;
Agarwal et al. 2012; Fernandez et al. 2014; Ferrara et al. 2014; Visbal et al. 2014b) and cosmological
simulations (Agarwal et al. 2014) suggest that such fields may have
been produced regularly in the early universe and therefore that DCBH formation may have 
been relatively common.\footnote{Recently, Dijkstra et al. (2014) have
  investigated the dependence of this conclusion on input parameters
  that are often assumed in modeling and simulations.}  This conclusion has been strengthened by complementary
modeling of the growth (e.g. Johnson et al. 2012; Latif et al. 2013b;
Regan et al. 2013) and evolution
(e.g. Begelman 2010; Hosokawa et al. 2012, 2013; Inayoshi et al. 2013;
Schleicher et al. 2013) of the supermassive stellar progenitors of
DCBHs, which suggests that their growth to mass scales of up to $\sim$
10$^6$ M$_{\odot}$ is not likely impeded by radiative feedback or
pulsational instability.  Indeed, there is a strong possibility that
DCBH remnants still reside in present-day galaxies (Koushiappas et
al. 2004; Bellovary et
al. 2011; Devecchi et al. 2012; Greene 2012; Reines et al. 2014), including our
Milky Way (Rashkov \& Madau 2014) and its satellites (van Wassenhove et al. 2010)

While the elevated LW radiation fields required for widespread DCBH
formation may have been produced in the early Universe, it is likely
that in many cases other forms of radiation accompanied them.  In
particular, given that the sources of this radiation were likely early
star forming galaxies, ionizing radiation from stars, X-rays from
young accreting BHs and cosmic rays from early supernovae may have
commonly been present, in addition to LW radiation.  Using one zone
models of primordial gas collapse in atomically cooled haloes, 
Inayoshi \& Omukai (2011) addressed the role that X-rays and cosmic ray ionization
may have played in the process of DCBH formation, finding that the
free electron population generated deep in the cores of the haloes could 
have catalyzed the rapid formation of H$_{\rm 2}$ molecules, leading
to cooling of the primordial gas to temperatures well below 10$^3$ K and preventing DCBH formation.  
In addition, Yue et al. (2014) have recently developed a semi-analytic
model in which DCBH formation is halted once the Universe becomes reionized,
due to the photoevaporation of the gas from atomic cooling haloes.

Here, we address the effects of ionizing radiation on the
formation of DCBHs using cosmological simulations that track 
both the elevated LW radiation field required for DCBH
formation as well as an accompanying ionizing radiation field.
In Section 2 we describe the methodology used to model
the effects of background ionizing and LW radiation fields.  In
Section 3 we present the results of our simulations.  Finally, in
Section 4, we give a brief discussion of our results.

  \begin{figure*}
   \includegraphics[angle=-90,width=6.8in]{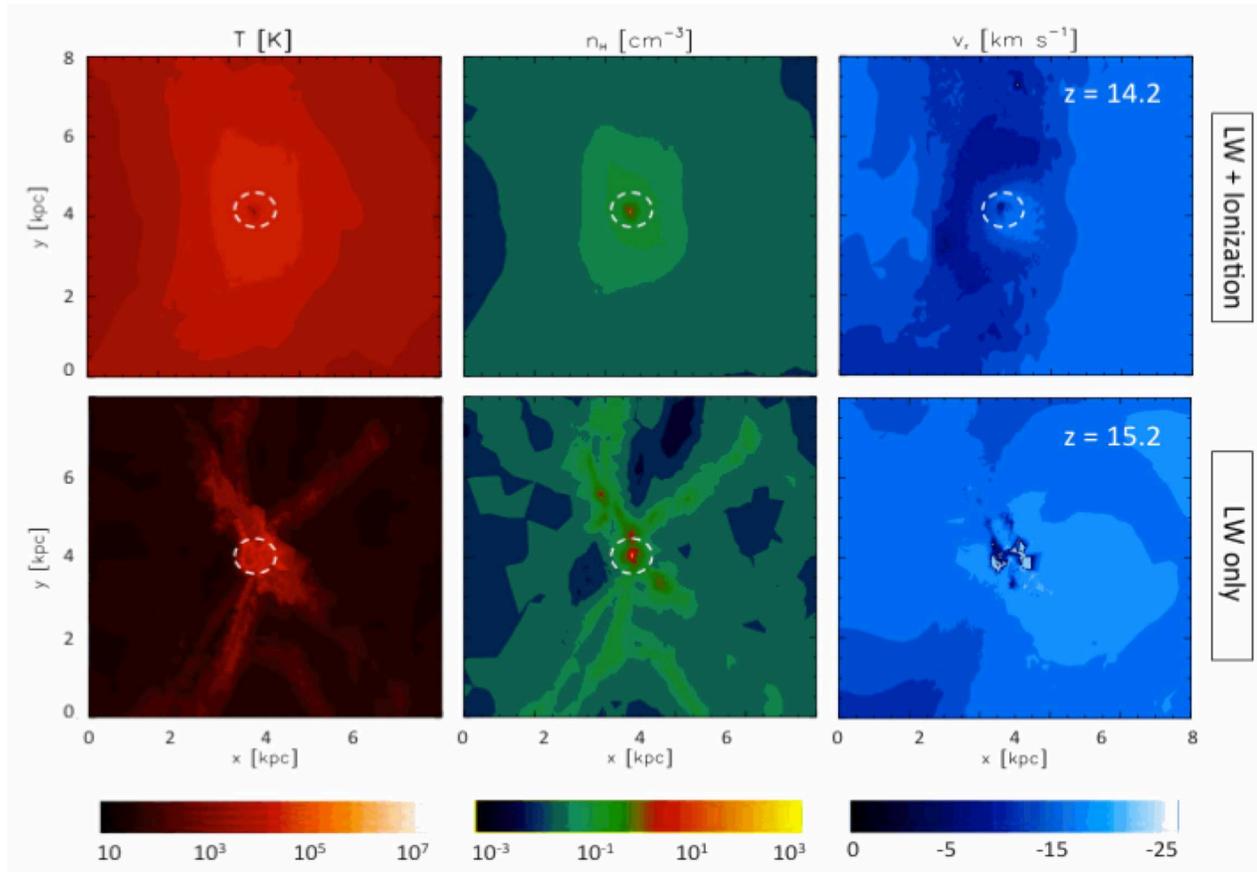}
   \caption{Properties of the primordial gas within a 10 pc
     (physical) slice of the simulation volume, centered on the
     atomic cooling halo, the location of the virial radius of which is
     denoted by the white dashed circles (but is suppressed in the
     bottom-right panel, for clarity).  From left to right, we show the
     temperature, number density of hydrogen atoms and the radial
     velocity if the gas relative to the center of the halo, for our
     simulation with LW and ionizing radiation included ({\it top})
     and for our simulation including only LW radiation ({\it
       bottom}).  The impact of the ionizing background radiation is
     to heat the gas in the IGM and to evaporate it out of the
     filaments that feed the central halo.  This
     results in slower infall of gas into the center of the halo, as
     evidenced by the smaller infall (negative) velocities in the case
with ionizing radiation included, as compared to the LW only
case.  This results in a delay of $\Delta$$z$ $\simeq$ 1 ($\simeq$ 25
Myr) in the onset of runaway gravitational collapse at the center of the halo.}
   \end{figure*}


\section{Methodology}
Here we describe the approach that we have taken to model the impact of a photoionizing background
on the process of DCBH formation.  We have carried out two cosmological simulations employing the same version of the smoothed particle hydrodynamics (SPH) code GADGET (Springel et al. 2001; Springel \& Herquist 2002) that we have employed in previous work (see e.g. Johnson et al. 2011, 2013).
We make use of the same initial conditions as in those previous works,
namely a 1 Mpc$^3$ (comoving) cosmological volume which is initialized
at $z$ = 100 and within which an atomic cooling halo is identified at
$z$ $\simeq$ 15, at which time its DM mass is $\simeq$ 4 $\times$
10$^7$ M$_{\odot}$ (corresponding roughly to a 3-$\sigma$ fluctuation).
It is the evolution of the primordial gas during its collapse into this halo that is the focus of our study. 

In both of these simulations we use the same prescription for the LW
background radiation field that we have employed in the previous works
cited above.  This consists of a constant uniform background LW field
with an intensity characterized by $J_{\rm 21}$ = 10$^3$, with a
spectrum charaterized by a temperature of $\sim$ 10$^4$ K (Shang et al. 2010).\footnote{Here we
  follow the standard convention and assign to J$_{\rm 21}$ units of
  10$^{-21}$ erg s$^{-1}$ cm$^{-2}$ Hz$^{-1}$ sr$^{-1}$.}  We account for the self-shielding of H$_{\rm 2}$ molecules to this radiation by calculating the H$_{\rm 2}$
self-shielding factor, which expresses the fraction of the
unattenuated background LW flux to which a gas parcel is exposed,
using an estimate based on the local column density of H$_{\rm 2}$
molecules (Bromm \& Loeb 2003; see also Shang et al. 2010; Wolcott-Green et al. 2011).

Along with this constant (in time) and uniform LW radiation field, we model the effect of an accompanying 
constant uniform ionizing radiation field.\footnote{While the LW
  background radiation is turned on at $z$ = 100 in our simulations,
  the ionizing background is turned on at $z$ = 30 in our fiducial
  case.  In Appendix B, we discuss the impact of turning the ionizing
  background on later, at $z$ = 20.}  We adopt the following value for the local
photoionization rate, assuming a ratio of ionizing photons to LW
photons appropriate for a low-metallicity stellar population with an age of $\sim$ 10$^7$ yr
(Leitherer et al. 1999):

\begin{equation}
\Gamma_{\rm ion} = 1.5 \times 10^{-11}  e^{-\tau_{\rm ion}} \, {\rm s}^{-1} \mbox{\ ,}
\end{equation}
where $\tau_{\rm ion}$ is the local optical depth to ionizing photons, estimated as 
described below.
The value this obtains for the unattenuated photoionization rate (with
$\tau_{\rm ion}$ = 0)
is chosen to be consistent with our choice of $J_{\rm 21}$ = 10$^3$ for
the unattenuated LW flux, assuming a value for the ratio of the escape
fraction of ionizing photons to the escape fraction of LW photons for
the galaxies producing the radiation of $f_{\rm esc,ion}$/$f_{\rm
  esc,LW}$ $\sim$ 0.3, which is in broad agreement with estimates
gleaned from simulations of early dwarf galaxies
(e.g. Ricotti et al. 2001; Kitayama et al. 2004; Wise \& Cen 2009;
Razoumov \& Sommer-Larsen 2010; Paardekooper et al. 2013).\footnote{We note 
that if a smaller fraction of ionizing photons relative to
LW photons escape from source galaxies or if the stellar population is
much older than $\sim$ 10$^7$ yr, then the photoionization 
rate we have adopted will be an overestimate.  In particular, the rate,
normalized to the LW flux, may be roughly two
orders of magnitude lower for a population age of $\sim$ 10$^8$ yr
(e.g. Leitherer et al. 1999), although we emphasize that in the early
universe (e.g. at $z$ $\ga$ 6) the stars producing the bulk of the LW flux are likely much younger
than this.}
For the corresponding photoheating rate, we conservatively assume that
$\simeq$ 2 eV is deposited in the gas as heat for each
photoionization. 

We adopt a local approximation for the flux of ionizing photons to
which a given parcel of gas is exposed, by estimating the optical
depth to ionizing photons as

\begin{equation}
\tau_{\rm ion} = \sigma_{\rm ion} n_{\rm H} r_{\rm char} \simeq 10^2 \, \left(\frac{h\nu}{13.6 \, {\rm eV}} \right)^{-3} \left(\frac{n_{\rm H}}{1 \, {\rm cm}^{-3}} \right) \left(\frac{r_{\rm char}}{5 \, {\rm pc}} \right) \mbox{\ ,}
\end{equation}
where $\sigma_{\rm ion}$ $\simeq$ 6 $\times$ 10$^{-18}$ cm$^{-2}$
($h\nu$/13.6 eV)$^{-3}$ is the cross section for photoionization of
neutral hydrogen,\footnote{We evaluate this cross section at $h\nu$ = 15.6 eV,
consistent with the 2 eV that we assume is deposited in the gas for each
photoionization.} $n_{\rm H}$ is the local density of neutral hydrogen
atoms and $r_{\rm char}$ is the physical length scale appropriate for the
parcel of gas, which we take to be defined in terms of the mass
$m_{\rm SPH}$=120 M$_{\odot}$ of an SPH particle in our simulation:

\begin{equation}
r_{\rm char} = \left(\frac{3}{4 \pi} \frac{m_{\rm SPH}
  }{\rho}\right)^{\frac{1}{3}} \simeq 10 \,  \left( \frac{n_{\rm H}}{1 \, {\rm cm^{-3}}} \right)^{-\frac{1}{3}} \, {\rm pc} \mbox{\ ,}
\end{equation}
 where $\rho$ is the gas density at the location of the SPH particle.
Note that we have normalized equations (2) and (3) for a neutral 
primordial gas; in particular, the optical depth can be 
$\tau_{\rm ion}$ $<<$ 1 where the gas is highly ionized.

While this method provides a simple approximation for the local
ionization and heating rates, as discussed further in Appendix A, we expect that our local estimate
for the optical depth to ionizing photons (equation 2) is valid. This
is supported by the fact that the results of our
simulations for the gas density and collapse redshift required for
self-shielding from ionizing radiation are in very good agreement with
the detailed estimates recently presented by Noh \& McQuinn
(2014; see their equations 6 and 7).  Our results are also consistent with those found by
Dijkstra et al. (2004) for the conditions required for the gas in high 
redshift haloes to self-shield against an intergalactic photoionizing 
background.  These authors found that the gas self-shields 
more readily at high redshift due to the higher densities in
virialized haloes.  The atomic cooling halo we focus on in our 
simulations, and into which gas does collapse in the presence of an
ionizing backgroud, has a circular velocity of  $\simeq$ 20 km s$^{-1}$ 
at $z$ $\simeq$ 15, which is higher than the minimum circular velocity 
that Dijkstra et al. (2004) find is required to retain gas in the
presence of a photoionizing background at this redshift.  Furthermore,
this is also consistent with the results of Okamoto et al. (2008), who found from
cosmological simulations that haloes with circular velocities
$\ga$ 10 km s$^{-1}$ are able to retain a large fraction of their
baryonic mass at $z$ $\ga$ 10.
This agreement with previous work gives us confidence that our approach 
produces reliable results.

  \begin{figure}
    \includegraphics[width=3.3in]{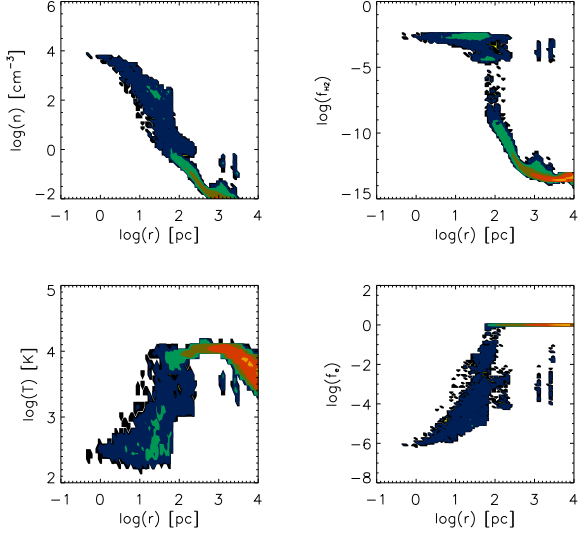}
    \caption{Properties of the primordial gas in our simulation
      inluding both LW and ionizing radiation backgrounds, when the
      number density of hydrogen atoms in the center of the host halo has reached
      $n$ $\simeq$ 10$^4$ cm$^{-3}$, at the same time as shown in the
      top panel of Figure 1 ($z$ = 14.2).  Clockwise from top-left:  the number density
      of hydrogen atoms, the H$_{\rm 2}$ fraction, the free electron
      fraction and the temperature.  Contours denote the distribution
      of the gas, with the mass fraction varying by an order of
      magnitude across contour lines.  The high free electron fraction
      in the photoionized gas leads
      to efficient H$_{\rm 2}$ formation within the
      central $\sim$ 100 pc where the gas is self-shielded to the
      ionizing radiation.  In turn, this leads to efficient cooling of
    the gas by H$_{\rm 2}$, despite the elevated molecule-dissociating
    LW radiation field, and the temperature falls to 200 K.  
    It is unlikely that a DCBH can form from gas at such low temperatures.}
  \end{figure}

  \begin{figure}
    \includegraphics[width=3.3in]{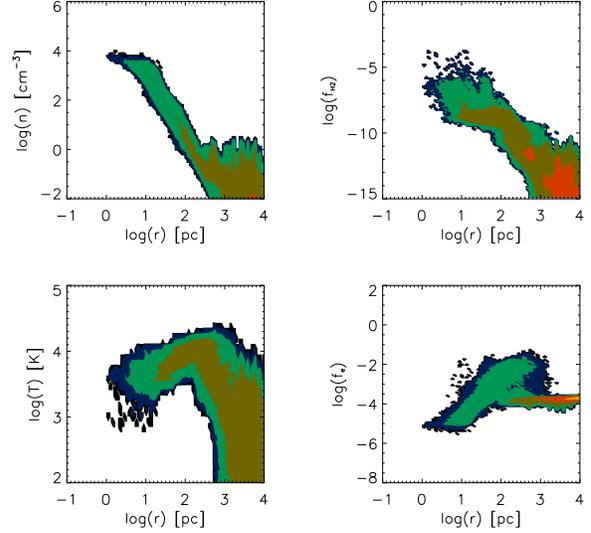}
    \caption{The same as Figure 2, but for our simulation including
      only a LW background radiation field, at the same time as shown in the
      bottom panel of Figure 1 ($z$ = 15.2).  In this case, the free electron fraction in the gas
      passsing through the virial shock never becomes larger than $\sim$ 10$^{-2}$ and,
      consequently, the H$_{\rm 2}$ fraction remains below $\sim$
      10$^{-5}$.  At such a low abundance H$_{\rm 2}$ cooling does
      not operate efficiently and the gas stays at temperatures high
      enough for a DCBH to form.}
  \end{figure}

\section{Results}
Here we compare the results of our two simulations, highlighting the effects of the ionizing 
background on the evolution of the primordial gas.  As we shall see, this radiation 
can profoundly alter the final outcome of the collapse of the gas.

Figure 1 shows the properties of the gas within a 10 pc (physical)
slice of the simulation volume, centered on the atomic cooling halo,
when the gas at the center of the halo has collapsed to a maximum
density of $n_{\rm H}$ $\simeq$ 10$^4$ cm$^{-3}$.  This corresponds to
redshifts $z$ = 14.2 and 15.2, for the simulations with and without
ionizing radiation, respectively.  The effect of the photoheating in
the low density intergalactic medium (IGM) is evident in the top
panels, which show that the gas is evaporated out of the cosmological
filaments that feed the central halo; as shown in the bottom panels,
these structures are intact in the simulation with only LW radiation.
This photoheating and evaporation of the filamentary gas results in a
higher gas pressure and a slower accretion rate of gas into the halo, as is shown by the lower (negative) infall velocities of the gas in the panels on the right.  This results in the gas collapsing to $n_{\rm H}$ $\simeq$ 10$^4$ cm$^{-3}$ roughly 25 Myr later in the simulation including ionizing radiation.

Figures 2 and 3 show the properties of the gas in the central halo, as functions of the distance from the densest gas particle.  In contrast to the case with just LW radiation, in the case with ionizing radiation, at large distances the gas is highly ionized and is almost entirely heated to 
temperatures up to $\simeq$ 10$^4$ K.  At distances $\la$ 100 pc, the
gas is self-shielded to the ionizing radiation and can collapse
to high densities.  Due to the higher free electron fraction of the collapsing gas in the case with ionizing radiation, at high densities the formation of H$_{\rm 2}$ is catalyzed (via the same reactions through which H$_{\rm 2}$ is generally formed in the primordial gas; see e.g. Bromm \& Larson 2004) and a much larger fraction of H$_{\rm 2}$ is generated than in the case with just LW radiation. 

Due to the higher H$_{\rm 2}$ fraction in the case with ionizing
radiation, the gas is able to cool via molecular transitions down to
$\sim$ 200 K, as is typical for Population (Pop) III star formation in
minihaloes (see e.g. Greif et al. 2011; Clark et al. 2011).  These
temperatures are much lower than the $\sim$ 10$^3$ -- 10$^4$ K to
which the gas can cool in the case with LW radiation only.  The
corresponding accretion rates onto the objects that form via the
runaway gravitational collapse of the gas at the center of the halo
are expected to  be very different, because of the large difference in the gas temperature (e.g. Omukai \& Palla 2003):

\begin{equation}
\frac{dM_{\rm acc}}{dt} \simeq 10^{-3} \, \left(\frac{T}{200 \, {\rm K}} \right)^{\frac{3}{2}} {\rm M_{\odot}} \, {\rm yr^{-1}}  \mbox{\ .}
\end{equation}
Using the temperatures at the center of the halo shown in Figs. 2 and
3, for the cases with and without ionizing radiation, this corresponds
roughly to $\simeq$ 10$^{-3}$ M$_{\odot}$ yr$^{-1}$ and $\simeq$
10$^{-1}$ M$_{\odot}$ yr$^{-1}$, respectively.  Again, the former is
consistent with Pop~III star formation in which H$_{\rm 2}$ cooling is
effective, while the latter is consistent instead with the formation
of a supermassive star (or binary supermassive stars; see e.g. Regan
\& Haehnelt 2009; Whalen et al. 2013) with a mass of $\sim$ 10$^5$
M$_{\odot}$ (e.g. Wise et al. 2008; Shang et al. 2010; Johnson et
al. 2012; Latif et al. 2013b).  From this, we can conclude that when
there is only LW radiation it is likely that a supermassive star (and
subsequently a DCBH) will form, while in the case with both LW and
ionizing radiation a small galaxy composed of Pop~III stars with masses $\la$ 10$^3$ M$_{\odot}$ will likely form instead (e.g. Hirano et al. 2014).  

Figure 4 shows the enclosed mass of H$_{\rm 2}$, as a function of
distance from the center of the halo, in both simulations.  The much 
higher H$_{\rm 2}$ mass in the case with ionizing radiation, leads to 
strong self-shielding of the gas to LW radiation, as shown in  
Figure 5.  This effective self-shielding in turn implies that the
H$_{\rm 2}$ photodissociation rate in the center of the halo is much
lower than in the case with just LW radiation.  This leads to a higher
H$_{\rm 2}$ fraction, which in turn leads to stronger self-shielding
of the gas.  Thus, there is a runaway process of H$_{\rm 2}$ formation
and self-shielding, set up by the fact that the gas collapses into the
halo with an elevated free electron fraction because it was photoionized in the IGM.  The slower formation of H$_{\rm 2}$ in the case with just LW radiation leads to weaker self-shielding and, ultimately, to the weak molecular cooling and higher temperatures that set the stage for the formation of a DCBH.

Figure 6 shows both the mass enclosed and the radially-averaged Jeans
mass of the gas, as a function of the distance from the center of the
halo, for both simulations.  Runaway
gravitational collapse is possible when the enclosed mass
exceeds the Jeans mass.  As shown in Figures 6 and 7, there is a
factor of a few less gas in the center of the halo in the case with ionizing
radiation.  However, due to the lower temperatures of the gas in this
case, we expect that the central $\la$ 10$^4$ M$_{\odot}$ of gas
becomes Jeans unstable and undergoes runaway gravitational collapse,
while in the case with just LW radiation the gas is Jeans unstable at
a much larger mass scale of $\sim$ 10$^5$ M$_{\odot}$.  This is again
consistent with our expectation that a supermassive star, and
subsequently a DCBH, will form in the case with just LW radiation,
while a small Pop~III galaxy will instead form in the case including ionizing radiation. 

   \begin{figure}
    \includegraphics[width=3.4in]{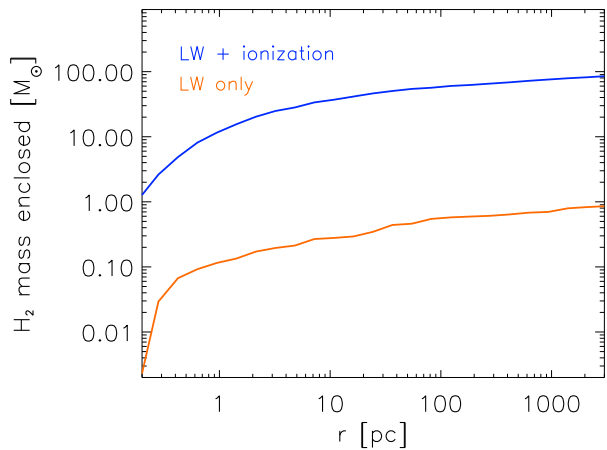}
    \caption{The mass in H$_{\rm 2}$ molecules enclosed, as a function of distance from the center of the atomic cooling halo, in our two simulations at the same redshifts as shown in the previous Figures.  Due to the elevated free electron fraction in the case with photoionization, a larger mass of H$_{\rm 2}$ builds up than in the case with just LW radiation.  This leads to the gas becoming self-shielding to LW radiation, as shown in Figure 5.}
  \end{figure}

   \begin{figure}
    \includegraphics[width=3.4in]{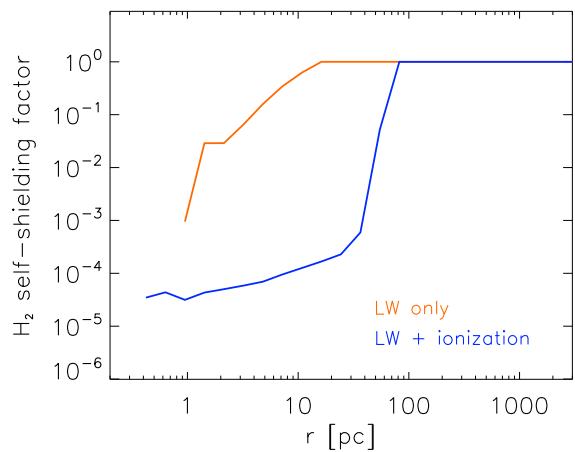}
    \caption{The factor by which the H$_{\rm 2}$-dissociating LW flux
      is decreased due to local self-shielding of the H$_{\rm 2}$
      molecules, as a function of distance from the center of the
      host halo.  Shown are the radially-averaged values of this
      factor, for LW and photoionizing
      background radiation fields ({\it blue}) and for just a LW
      background radiation field ({\it orange}), at the same times as
      shown in the previous Figures.  The elevated H$_{\rm 2}$
      fraction in the simulation including ionizing radiation leads to
    much stronger self-shielding of the molecules, which allows for
    efficient formation of H$_{\rm 2}$.  }
  \end{figure}

   \begin{figure}
    \includegraphics[width=3.4in]{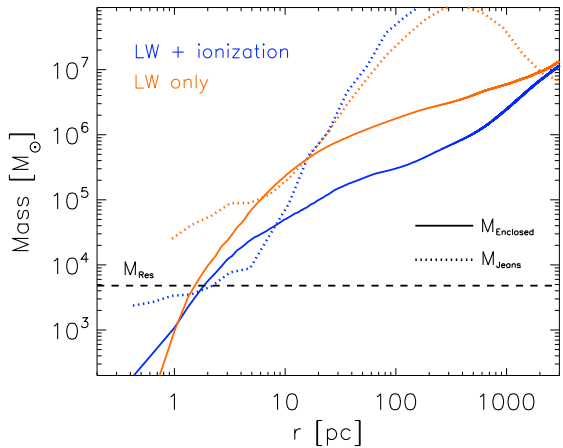}
    \caption{The mass enclosed ({\it solid lines}) and the
      radially-averaged Jeans mass ({\it dotted lines}), as functions
      of the distance from the center of the host atomic cooling halo,
      with LW and photoionizing
      background radiation fields ({\it blue}) and with just a LW background radiation field ({\it orange}),
      at the same times as shown in previous Figures.
Also shown is the mass contained in the SPH
      smoothing kernel in our simulations, $m_{\rm Res}$, which
      roughly corresponds to the minimum mass that can be
      resolved.  The gas is only able to collapse under its own gravity when the
      enclosed mass is larger than the Jeans mass.  In the case with
      LW and ionization this occurs at a mass scale of
      a few $\times$ 10$^3$ M$_{\odot}$, whereas in the case with only LW radiation
     this occurs at a mass scale of $\sim$ 10$^5$
      M$_{\odot}$.  In the former case, the mass scale is too small
      for the formation of a DCBH, whereas it is sufficiently large 
    for a DCBH to form in the latter.}
  \end{figure}

   \begin{figure}
    \includegraphics[width=3.4in]{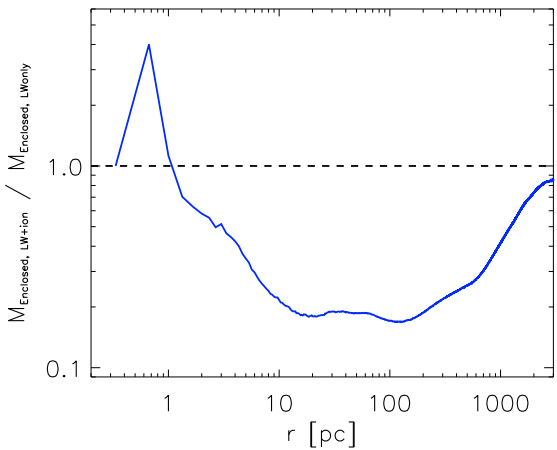}
    \caption{The ratio of the mass enclosed in the simulation
      including both LW and ionizing radiation to that including only
      LW radiation, as a function of distance from the center of the
      host atomic cooling halo.  Due to the higher pressure of the gas
    in the simulation including ionizing radiation, the accretion rate
  into the center of the halo is lower than in the case without it and
the mass enclosed within the virial radius at $r$ $\simeq$ 500 pc is
lower by a factor of $\simeq$ 3.
Note that, as shown in Figure 6, the mass within $r$ $\simeq$ 2 pc is
not well-resolved in our simulation.}
  \end{figure}

\section{Discussion}
We have presented a pair of cosmological simulations which demonstrate that it is 
possible for ionizing radiation 
to prevent DCBH formation in atomic cooling haloes.  We find that this
is due to the large free electron fraction in
the photoionized gas, which leads to the rapid formation of H$_{\rm
  2}$ and ultimately to effective
molecular cooling that sterilizes the halo for DCBH formation.    
We expect that a small Pop~III galaxy with up to $\sim$ 10$^4$ M$_{\odot}$ in primordial
stars may form, instead, similar to previously studied cases of
Pop~III star formation in atomic cooling haloes including just a background
LW radiation field (e.g. Oh \& Haiman 2002; Trenti \& Stiavelli 2009; 
Safranek-Shrader et al. 2012).

We emphasize that this effect relies on the gas being photoionized before its collapse into the 
center of the halo.  In cases where the ionizing radiation turns on at relatively
late times, when the gas in the halo has already collapsed to densities at which it
is self-shielded from the ionizing radiation, we do not expect that a large H$_{\rm 2}$ 
fraction develops, and indeed it is likely that DCBH formation is still able
to take place in this case.  In Appendix B, we present one such
case, in which the ionizing background turns on at $z$  = 20, by which
time the dense gas in the halo is already self-shielded from the
ionizing radiation.  Indeed, we find that in this case the collapse of
the self-shielded gas proceeds almost identically to the case with
just LW radiation (compare Figures 3 and B1).  Because many atomic cooling haloes may be 
self-shielding to ionizing radiation, especially in the early stages of
reionization (see e.g. Noh \& McQuinn 2014), DCBH
formation may still take place readily at $z$ $\ga$ 15 even in photoionized
regions.  

We note that this picture is in basic agreement with that presented by Yue et
al. (2014), who argue that in photoionized regions the gas can not be
retained by atomic cooling haloes at $z$ $\la$ 14, and so DCBHs will not likely
form in photoionized regions below this redshift.  Our result is
related, but distinct -- we find that if the gas in atomically
cooling haloes is subject to an elevated ionizing radiation background
and can not self-shield against it, then DCBH formation can be
prevented {\it even if the gas is retained in the halo}.  We also emphasize that, because
reionization is an inhomogeneous process, it should in principle be possible
for DCBHs to form at redshifts $z$ $\la$ 14 in regions which are not yet
reionized; consistent with this, Agarwal et al. (2014) find that DCBHs 
may form down to at least $z$ $\sim$ 9.

Our results suggest that, when ionizing radiation accompanies a
background LW radiation field the critical LW flux required for DCBH 
formation is likely to be significantly higher than in the absence of 
ionizing radiation.  As we adopted $J_{\rm 21}$ = 10$^3$ for our
simulations, it appears that the critical flux for photoionized gas
may be well above this value, although this is likely to vary with
redshift and may be a function of the growth history of the halo (see e.g. Latif et al. 2014).
This correction could lead to reduced estimates of the 
prevalence of DCBHs in the early Universe, as compared
to previous results (see e.g. Dijkstra et al. 2008; Shang et al. 2010; 
Agarwal et al. 2012, 2014; Latif et al. 2014), at least in reionized
regions.  

Related to the question of the critical LW flux, we note that here we 
have followed the approach of similar works (e.g. Bromm \& Loeb 2003;
Shang et al. 2010) and adopted the approximation given by Draine \& Bertoldi (1996)
for the H$_{\rm 2}$ self-shielding factor.  As shown by Shang et
al. (2010) and Wolcott-Green et al. (2011), this likely overestimates
the shielding factor, and so may lead to overestimates of the critical
LW flux required to suppress molecular cooling (but see also Richings
et al. 2014).  Therefore, the critical flux for the halo in our
simulations may in fact be significantly lower than the $J_{\rm 21}$ = 10$^3$ that
we adopted.  As we have used the same self-shielding prescription
in our simulations both with and without ionizing radiation, we do not
expect that using an improved prescription would qualitatively change
our central results pertaining to the effects of ionizing radiation.  
That said, it may be crucial to model in great detail
both the photodissociation of H$_{\rm 2}$ molecules and the
transfer of the H$_{\rm 2}$ line emission that can cool the gas
(e.g. Greif 2014), in order to determine the final fates of atomic
cooling haloes exposed to elevated levels of LW radiation.

We expect that the Pop~III galaxies that form in cases in which ionizing radiation
prevents DCBH formation may contain significantly more mass in
primordial stars than the stellar clusters formed
in smaller minihaloes.  This follows from 
the fact that the relatively deep DM gravitational potential wells of atomic cooling haloes 
allow the gas to be more readily retained in the face of stellar feedback than in the
case of Pop~III star formation in minihaloes (see e.g. Bromm \& Yoshida 2011).
Due to their higher masses, we expect that Pop~III galaxies formed in atomic cooling haloes
may be among the most luminous Pop~III star-forming objects, with
distinct observational signatures that could be detected by future missions such as the {\it James Webb Space Telescope} (see e.g. Schaerer 2003; Johnson 2010; Inoue 2011; Zackrisson et al. 2011; Pawlik et al. 2013).

\section*{Acknowledgements}
Work at LANL was done under the auspices of the National Nuclear Security 
Administration of the US Department of Energy at Los Alamos National 
Laboratory under Contract No. DE-AC52-06NA25396.  All simulations were
performed on the Institutional Computing (IC) network platform Mustang
at LANL.  D.J.W. was supported by the European Research Council under the European Community's 
Seventh Framework Programme (FP7/2007 - 2013) via the ERC Advanced Grant "STARLIGHT:  
Formation of the First Stars" (project number 339177).  We would like to thank Joseph Smidt, John Wise and Hui Li
for valuable discussions which helped to shape this work.


\appendix

\section{Local approximation for the photoionization rate}
Our approach to modelling the impact of an ionizing background 
radiation field, as described in Section 2, relies on a simple,
local estimate of its attenuation.  In particular, it is assumed
that the attenuation of the field largely takes place over 
a length-scale comparable to, or smaller than, that of an 
individual SPH particle.  Here we show that this
assumption is valid.

The optical depth to ionizing radiation over the length-scale
of a parcel of gas represented by a single SPH particle, under the
assumption of a neutral medium,
is given by equation (2).   Evaluated at a photon energy of 
15.6 eV (as we assume in our calculations; see Section 2)  
and expressing this length-scale following equation (3), 
the optical depth becomes
  
\begin{equation}
\tau_{\rm ion} \simeq 1.3 \times 10^2 \,  \left(\frac{n_{\rm H}}{1 \, {\rm cm}^{-3}} \right)^{\frac{2}{3}}  \mbox{\ .}
\end{equation}
As shown in Figures 2 and 3, the density profile of the atomic
cooling halo in our simulations can be approximated as 
isothermal, with the number density of hydrogen nuclei
approximated as $n_{\rm H}$ $\simeq$ 10$^5$ ($r$/pc)$^{-2}$.  
Using this expression in the above formula for 
optical depth, we obtain

\begin{equation}
\tau_{\rm ion}   \simeq 3 \times 10^5 \,  \left(\frac{r}{{\rm pc}} \right)^{-\frac{4}{3}} \mbox{\ ,}
\end{equation} 
where here $r$ is the distance from the center of the halo.
This implies that in neutral regions the optical depth is large, i.e. that $\tau_{\rm ion}$ $\ga$ 1, 
over the length-scale of an SPH particle at $r$ $\la$ 10$^4$ pc.
Therefore, we expect our approximation to be valid in the dense central 
regions of the halo, at $r$ $\sim$ 100 pc, within which we find the
gas to be optically thick to the external ionizing radiation field.

\section{Evolution of self-shielded gas}
Here we briefly highlight the results of a simulation in which
the ionizing background turns on at $z$ = 20, in order to 
highlight the dependence of our results on the state of the halo 
when first subjected to the ionizing background.  By this redshift, the
dense gas in the halo is already self-shielding to the ionizing 
radiation and it is never photoionized.  We note that the gas becoming
self-shielded already by $z$ = 20 in this halo is consistent with the 
recent analytical estimates presented by Noh \& McQuinn (2014), as
well as with the results of previous numerical simulations cited in 
Section 2 (Dijkstra et al. 2004; Okamoto et al. 2008).

Figure B1 shows the properties of the gas, as functions of the distance from the 
densest SPH particle in the halo, just before it undergoes 
runaway gravitational collapse.  Comparing this to Figure 3,
we see that the gas evolves in much the same way as in the case with
just LW radiation.  In particular, the free electron fraction and the H$_{\rm 2}$
fraction in the inner regions of the halo are comparably low in both
cases, and the gas remains sufficiently hot for DCBH formation to
proceed.  This supports our conclusion that DCBH formation is
prevented by ionizing radiation only in cases in which the dense 
gas in the core of the halo is not self-shielding to the radiation and
undergoes runaway gravitational collapse only after being photoionized.

  \begin{figure}
    \includegraphics[width=3.3in]{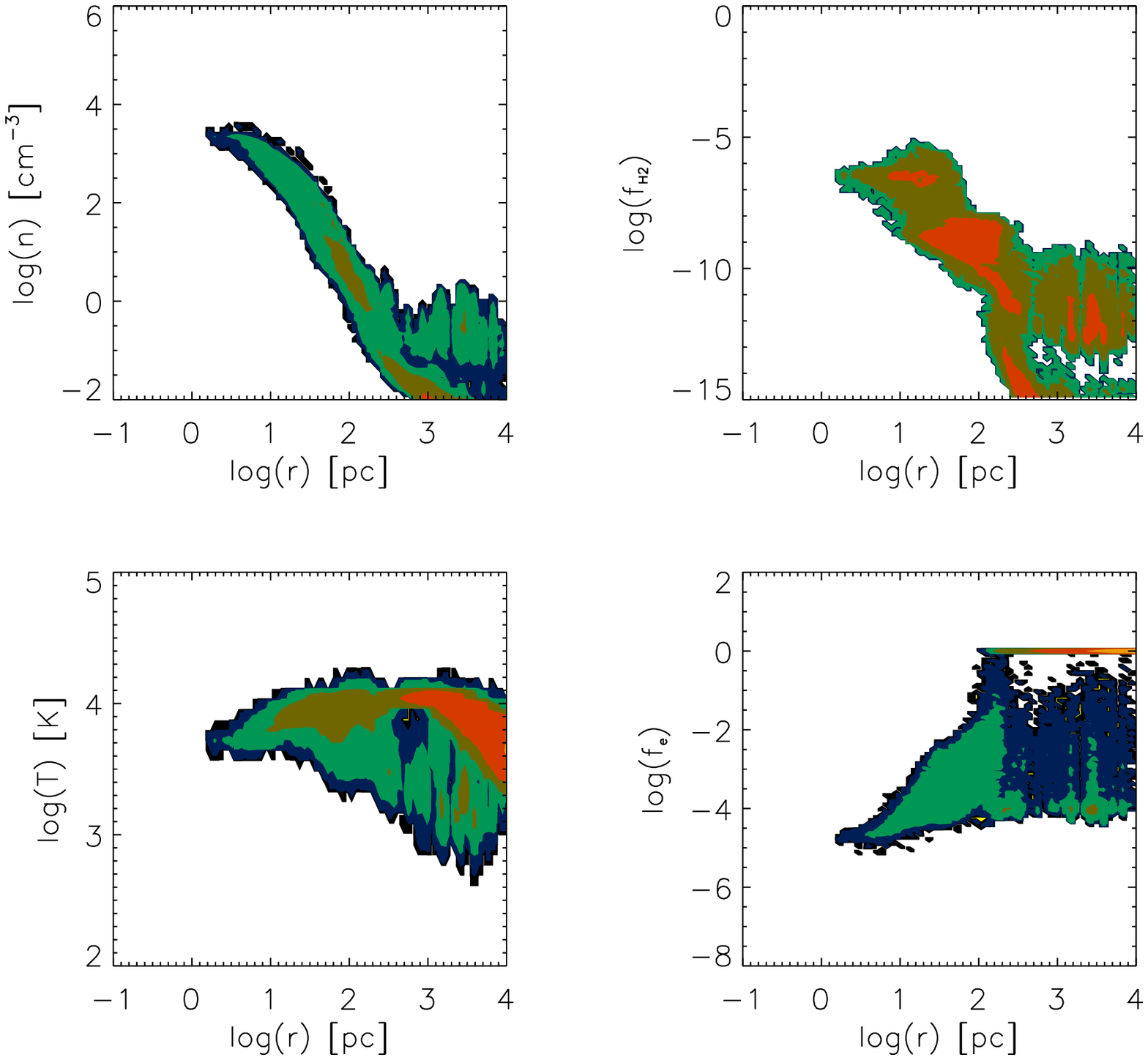}
    \caption{The same as Figures 2 and 3, but for our simulation in which
      the photoionizing background is turned on at $z$ = 20.  In this
case, the gas in the halo has already collapsed to relatively high 
density by this redshift and the dense core of the halo is
self-shielding to the ionizing radiation.  As a result, the free 
electron fraction in the dense gas is not elevated relative to the
case with just LW radiation and the H$_{\rm 2}$ fraction also remains
low (compare to Figure 3).  Therefore, in this case and in others
where the dense gas in the core of the halo is self-shielded to the
ionizing radiation and never becomes photoionized, we expect that
DCBH formation is not prevented and likely proceeds just as in the
case with only LW radiation.}
  \end{figure}


\begin{thebibliography}{99}

\bibitem[2(2000)]{b}Agarwal, B., Khochfar, S., Johnson, J.~L., Neistein, E., Dalla Vecchia, C., Livio, M. 2012, MNRAS, 425, 2854
\bibitem[2(2000)]{b}Agarwal, B., Dalla Vecchia, C., Johnson, J.~L., Khochfar, S., Paardekooper, J.-P. 2014, MNRAS, submitted (arXiv:1403.5267)
\bibitem[2(2000)]{b}Ahn, K., Iliev, I., Shapiro, P.~R., Mellema, G.,
  Koda, J., Mao, Y 2012, ApJ, 756, 16
\bibitem[2(2000)]{}Begelman M.~C. 2010, MNRAS, 402, 673
\bibitem[2(2000)]{}Begelman M.~C., Rossi, E.~M., Armitage, P.~J. 2008,
  MNRAS, 370, 289
\bibitem[2(2000)]{}Bellovary, J., Volonteri, M., Governato, F., Shen,
  S., Quinn, T., Wadsley, J. 2011, ApJ, 742, 13
\bibitem[2(2000)]{b}Bromm, V., Loeb, A. 2003, ApJ, 596, 34
\bibitem[2(2000)]{b}Bromm, V., Yoshida, N. 2011, ARA\&A, 49, 373
\bibitem[2(2000)]{b}Bromm, V., Larson, R.~B. 2004, ARA\&A, 42, 79
\bibitem[2(2000)]{}Chen, K.-J., Heger, A., Woosley, S., Almgren, A., Whalen, D.~J., Johnson, J.~L. 2014, ApJ, submitted (arXiv:1402.4777)
\bibitem[2(2000)]{}Choi, J.-H., Shlosman, I., Begelman, M.~C. 2013, ApJ, submitted (arXiv:1304.1369)
\bibitem[2(2000)]{}Clark, P.~C., Glover, S.~C.~O., Smith, R.~J., Greift, T.~H., Klessen, R.~S., Bromm, V. 2011, Sci, 331, 1040
\bibitem[2(2000)]{b}Devecchi, B., Volonteri, M., Rossi, E.~M., Colpi,
  M., Zwart, S.~P. 2012, MNRAS, 421, 1465
\bibitem[2(2000)]{b}Dijkstra, M., Haiman, Z., Rees, M.~J., Weinberg, D.~H. 2004, ApJ, 601, 666
\bibitem[2(2000)]{b}Dijkstra, M., Haiman, Z., Mesinger, A., Wyithe, J.~S.~B. 2008, MNRAS, 391, 1961
\bibitem[2(2000)]{b}Dijkstra, M., Ferrara, A., Mesinger, A. 2014, MNRAS, 442, 2036
\bibitem[2(2000)]{b}Draine, B.~T., Bertoldi, F. 1996, ApJ, 468, 269
\bibitem[2(2000)]{}Fan, X., et al. 2006, AJ, 131, 1203 
\bibitem[2(2000)]{}Fernandez, R., Bryan, G.~L., Haiman, Z., Li,
  M. 2014, MNRAS, submitted (arXiv:1401.5803)
\bibitem[2(2000)]{}Ferrara, A., Salvadori, S., Yue, B., Schleicher, D.~R.~G. 2014, MNRAS, submitted (arXiv:1406.6685)
\bibitem[2(2000)]{}Fryer, C.~L., Heger, A. 2011, AN, 332, 408
\bibitem[2(2000)]{}Fuller, G.~M., Woosley, S.~E., Weaver, T.~A. 1986, 307, 675
\bibitem[2(2000)]{}Gebhardt, K., et al. 2000, ApJ, 539, L13
\bibitem[2(2000)]{}Glover, S.~C.~O., Brand, P.~W.~J.~L. 2001, MNRAS, 321, 385
\bibitem[2(2000)]{}Greene, J.~E. 2012, Nat, in press (arXiv:1211.7082)
\bibitem[2(2000)]{}Greif T.~H. 2014, MNRAS, submitted (arXiv:1404.5311)
\bibitem[2(2000)]{}Greif T.~H., Springel, V., White, S.~D.~M., Glover,
  S.~C.~O., Clark, P.~C., Smith, R.~J., Klessen, R.~S., Bromm,
  V. 2011, ApJ, 737, 75
\bibitem[2(2000)]{b}Haiman, Z., Abel, T., Rees, M.~J. 2000, ApJ, 534, 11
\bibitem[2(2000)]{b}Hirano, S., Hosokawa, T., Yoshida, N., Umeda, H., Omukai, K., Chiaki, G., Yorke, H.~W. 2014, ApJ, 781, 60
\bibitem[2(2000)]{b}Hosokawa, T., Omukai, K., Yorke, H.~W. 2012, ApJ,
  submitted (arXiv:1203.2613)
\bibitem[2(2000)]{b}Hosokawa, T., Yorke, H.~W., Inayoshi, K., Omukai,
  K., Yoshida, N. 2013, ApJ, submitted (arXiv:1308.4457)
\bibitem[2(2000)]{b}Inayoshi, K., Hosokawa, T., Omukai, K. 2013, MNRAS, accepted (arXiv:1302.6065)
\bibitem[2(2000)]{b}Inayoshi, K., Omukai, K. 2011, MNRAS, 416, 2748
\bibitem[2(2000)]{b}Inoue, A.~K. 2011, MNRAS, 415, 2920
\bibitem[2(2000)]{b}Johnson, J.~L., Khochfar, S., Greif, T.~H., Durier, F. 2011, MNRAS, 410, 919
\bibitem[2(2000)]{}Johnson, J.~L., Whalen, D.~J., Even, W., Fryer, C.~L., Heger, A., Smidt, J., Chen, K.-J. 2013, ApJ, 775, 107
\bibitem[2(2000)]{}Johnson, J.~L., Whalen, D.~J., Fryer, C.~L., Li, H. 2012, ApJ, 750, 66
\bibitem[2(2000)]{b}Johnson, J.~L. 2010,  MNRAS, 404, 1425
\bibitem[2(2000)]{b}Kitayama, T., Yoshida, N., Susa, H., Umemura, M. 2004, ApJ, 613, 631 
\bibitem[2(2000)]{}Koushiappas, S.~M., Bullock, J.~S., Dekel,
  A. 2004, MNRAS, 354, 292
\bibitem[2(2000)]{b}Latif, M.~A., Schleicher, D.~R.~G., Schmidt, W., Niemeyer, J. 2013a, MNRAS, submitted (arXiv:1304.0962)
\bibitem[2(2000)]{b}Latif, M.~A., Schleicher, D.~R.~G., Schmidt, W.,
  Niemeyer, J. 2013b, MNRAS, 436, 2989
\bibitem[2(2000)]{b}Latif, M.~A., Bovino, S., Van Borm, C., Grassi,
  T., Schleicher, D.~R.~G., Spaans, M. 2014, MNRAS, submitted (arXiv:1404.5773)
\bibitem[2(2000)]{b}Leitherer, C., et al. 1999, ApJS, 123, 3
\bibitem[2(2000)]{b}Lodato, G., Natarajan, P. 2006, MNRAS, 371, 1813
  Astropart. Phys., 31, 376
\bibitem[2(2000)]{b}Machacek, M.~E., Bryan, G.~L., Abel, T. 2001, ApJ, 548, 509
\bibitem[2(2000)]{}Merritt, D., Ferrarese, L. 2001, ApJ, 547, 140
\bibitem[2(2000)]{b}Mesinger, A., Bryan, G.~L., Haiman, Z. 2006, ApJ,
  648, 835
\bibitem[2(2000)]{}Montero, P.~J., Janka, H.-T., M{\" u}ller, E. 2012, ApJ, 749, 37
\bibitem[2(2000)]{}Mortlock, D.~J., et al. 2011, Nat, 474, 616
\bibitem[2(2000)]{b}Natarajan, P., Volonteri, M. 2012, MNRAS, 422, 2051
\bibitem[2(2000)]{b}Noh, Y., McQuinn, M. 2014, ApJ, submitted
  (arXiv:1401.0737)
\bibitem[2(2000)]{b}Oh, S.~P., Haiman, Z. 2002, ApJ, 569, 558
\bibitem[2(2000)]{b}Okamoto, T., Gao, L., Theuns, T. 2008, MNRAS,
  390, 920
\bibitem[2(2000)]{b}Omukai, K., Palla, F. 2003, ApJ, 589, 677
\bibitem[2(2000)]{b}Paardekooper, J.~P., Khochfar, S., Dalla Vecchia,
  C. 2013, MNRAS, 429, L94
\bibitem[2(2000)]{b}Pawlik, A.~H., Milosavljevi{\' c}, M., Bromm, V. 2013,
  ApJ, 767, 59
\bibitem[2(2000)]{b}Petri, A., Ferrara, A., Salvaterra, R. 2012,
  MNRAS, 422, 1690
\bibitem[2(2000)]{b}Prieto, J., Jimenez, R., Haiman, Z. 2013, MNRAS,
  436, 2301
\bibitem[2(2000)]{b}Rashkov, V., Madau, P. 2014, ApJ, 780, 187
\bibitem[2(2000)]{b}Razoumov, A.~O., Sommer-Larsen, J. 2010, ApJ, 710,
  1239
\bibitem[2(2000)]{}Regan J.~A., Haehnelt M.~G. 2009, MNRAS, 396, 343
\bibitem[2(2000)]{}Regan J.~A., Johansson, P.~H, Haehnelt M.~G. 2013,
  MNRAS, submitted (arXiv:1312.4962)
\bibitem[2(2000)]{}Reines, A.~E., Plotkin, R.~M., Russell, T.~D.,
  Mezcua, M., Condon, J.~J., Sivakoff, G.~R., Johnson, K.~E. 2014,
  ApJ, accepted (arXiv:1405.0278)
\bibitem[2(2000)]{}Richings, A.~J., Schaye, J., Oppenheimer,
  B.~D. 2014, MNRAS, 442, 2780
\bibitem[2(2000)]{b}Ricotti, M., Gnedin, N.~Y., Shull, M.~J. 2001,
  ApJ, 560, 580
\bibitem[2(2000)]{}Safranek-Shrader, C., Agarwal, M., Federrath, C.,
  Dubey, A., Milosavljevi{\' c}, M., Bromm, V. 2012, MNRAS, 426, 1159
\bibitem[2(2000)]{}Schaerer, D. 2003, A\&A, 397, 527
\bibitem[2(2000)]{}Schleicher, D.~R.~G., Palla, F., Ferrara, A.,
  Galli, D., Latif, M. 2013, A\&A, submitted (arXiv:1305.5923)
\bibitem[2(2000)]{b}Sethi, S., Haiman, Z., Pandey, K. 2010, ApJ, 721, 615
\bibitem[2(2000)]{b}Shang, C., Bryan, G.~L., Haiman, Z. 2010, MNRAS, 402, 1249
\bibitem[2(2000)]{b}Springel, V., Yoshida, N., White, S.~D.~M. 2001, NewA, 6, 79
\bibitem[2(2000)]{b}Springel, V., Hernquist, L. 2002, MNRAS, 333, 649
\bibitem[2(2000)]{b}Spaans, M., Silk, J. 2006, ApJ, 652, 902
\bibitem[2(2000)]{}van Wassenhove, S., Volonteri, M., Walker, M.~G.,
  Gair, J.~R. 2010, MNRAS, 408, 1139
\bibitem[2(2000)]{}Van Borm, C., Spaans, M. 2013, A\&A, submitted (arXiv:1304.4057)
\bibitem[2(2000)]{b}Visbal, E., Haiman, Z., Bryan, G.~L. 2014a, MNRAS, submitted (arXiv:1403.1293)
\bibitem[2(2000)]{b}Visbal, E., Haiman, Z., Bryan, G.~L. 2014b, MNRAS, submitted (arXiv:1406.7020)
\bibitem[2(2000)]{b}Volonteri, M. 2012, Sci, 337, 544
\bibitem[2(2000)]{}Volonteri, M., Begelman, M.~C. 2010, MNRAS, 409, 1022
\bibitem[2(2000)]{b}Whalen, D.~J., et al. 2013, ApJ, 778, 17
\bibitem[2(2000)]{b}Willott, C.~J., McLure, R.~J., Jarvis, M.~J. 2003,
  ApJ, 587, L15
\bibitem[2(2000)]{b}Wise, J.~H., Turk, M.~J., Abel, T. 2008, ApJ, 682, 745
\bibitem[2(2000)]{b}Wise, J.~H., Cen, R. 2009, ApJ, 693, 984
\bibitem[2(2000)]{}Wolcott-Green, J., Haiman, Z., Bryan, G.~L. 2011, MNRAS, 418, 838
\bibitem[2(2000)]{b}Yue, b., Ferrara, A., Salvaterra, R., Xu, Y., Chen, X. 2013, MNRAS, accepted (arXiv:1305.5177)
\bibitem[2(2000)]{b}Zackrisson, E., Inoue, A.~K., Rydberg, C.-E.,
  Duval, F. 2011, MNRAS, 418, L104

\end{thebibliography}
\end{document}